\documentstyle[twoside,fleqn,epsfig,espcrc2]{article}

\def\ea{{\em et al.}}

\def\CT{\cos\theta}
\def\C2T{\cos^2\theta}

\def\MZ{\rm M_Z}

\def\PIK{{\pi/\rm K}}

\def\PI0{{\pi^{\scriptscriptstyle{0}}}}
\def\PIPI0{\pi^\pm\PI0}

\def\A1{{\rm a_{\scriptscriptstyle{1}}}}

\def\PTAU{{\cal P}_\tau}

\def\STW{\sin^2\theta_W}

\def\AEL{{\cal A}_e}
\def\AL{{\cal A}_\ell}

\def\AT{{\cal A}_\tau}
\def\AU{{\cal A}_{e-\tau}}

\def\CTT{{\rm C_{TT}}}
\def\CTN{{\rm C_{TN}}}

\def\RAVL{{ \textstyle{2\,a_\ell\,v_\ell} \over \textstyle{|a_\ell|^2+|v_\ell|^2} }}

\def\EE{\rm e^+e^-}
\def\MM{\mu^+\mu^-}
\def\TT{\tau^+\tau^-}

\def\EEEE{\EE\rightarrow\EE}
\def\EEMM{\EE\rightarrow\MM}
\def\EETT{\EE\rightarrow\TT}
\def\EEFF{\EE\rightarrow\rm f\,\overline{f}\ }

\def\TAP{\tau\rightarrow\pi\nu }

\def\TAR{\tau\rightarrow\rho\nu }

\def\TA1{\tau\rightarrow\rm\!a_1\nu }

\def\TAX{\rm\tau\rightarrow\!X\nu^\prime s}

\title{Spin Analysis of the Process $\EETT$ at LEP}

\author{Pablo Garc\'{\i}a--Abia\address{DESY--IfH Zeuthen; Platanenallee 6;
        D--15738--Zeuthen, Germany}%
        \thanks{Invited talk to the 5${}^{th}$ Topical Seminar
        {\it The Irresistible Rise Of The Standard Model}, San Miniato (Italy),
        April 1997.}}
       
\begin{document}

\begin{abstract}  

Using  the  data  collected  by  the  four  experiments  at  LEP  during
1990--1994,   a  precise   measurement   of  the   $\tau$   longitudinal
polarisation ($\PTAU$) has been performed, as well as the measurement of
the  transverse--transverse  ($\CTT$)  and  transverse--normal  ($\CTN$)
$\tau$ spin correlations.  From the $\PTAU$ measurement, assuming lepton
universality  of the neutral  currents, the effective  weak mixing angle
has been  determined  to be $\STW = 0.2325  \pm  0.0006$.  The  Standard
Model predictions are consistent with the measured results.

\end{abstract}

\maketitle

\section{INTRODUCTION}

It is a well established principle that weak interactions violate parity.
This fact  constitutes  a powerful  tool for performing  high  precision
tests of the Standard Model~\cite{SM}.

At LEP, where $\EEFF$ ($f = e,\ \mu,\ \tau$)  interactions  are produced
at  centre of mass  energies  close to $\MZ$,  the study of  observables
related to the spin of the final state  fermions  provides  an  accurate
determination of the vector and  axial--vector  coupling of the fermions
to the Z boson.  Only in the  $\EETT$  processes,  where both taus decay
before  entering the detector, we can extract  information  on the final
state spin configuration by analysing the $\tau$ decay process itself.

The cross section describing the $\EETT$ process can be written in terms
of the final state taus spin components as follows~\cite{Berna}:
\begin{eqnarray}
\label{cross.section}
\frac{d^5\sigma}{ds_{\tau^{\pm}} d\Omega} & = &
\frac{1}{4}    |P(q^2)|^2\{{C_0}\,(1+\cos^2\theta)                                                 \\
       & +   & C_1\,2\cos\theta                                                          \nonumber \\
       & -   & D_0\,(s_{\tau^-}^L - s_{\tau^+}^L)\, (1+\cos^2\theta)                     \nonumber \\
       & -   & D_1\,(s_{\tau^-}^L - s_{\tau^+}^L)\, 2\cos\theta                          \nonumber \\
       & -   & C_0\, s_{\tau^-}^L s_{\tau^+}^L\, (1+\cos^2\theta)                        \nonumber \\
       & -   & C_1\, s_{\tau^-}^L s_{\tau^+}^L\, 2\cos\theta                             \nonumber \\
       & +   & C_2\,(s_{\tau^-}^Ns_{\tau^+}^N - s_{\tau^-}^Ts_{\tau^+}^T)\,\sin^2\theta  \nonumber \\
       & +   & D_2\,(s_{\tau^-}^Ns_{\tau^+}^T + s_{\tau^-}^Ts_{\tau^+}^N)\,\sin^2\theta\}\nonumber
\end{eqnarray}

\noindent  where $P(q^2)$  contains the Z propagator and $\theta$ is the
$\tau^-$  polar  angle  with  respect  to  the  incident   $e^-$  flight
direction.  $s_{\tau^-}^L$ is the (Longitudinal)  projection of the spin
of  the   $\tau$   along  its  flight   direction   in  the  lab  frame,
$s_{\tau^-}^T$  the (Transverse)  projection in the plane defined by the
incoming  $e^-$  and  the  outgoing  $\tau^-$  and  $s_{\tau^-}^N$   the
component Normal to these two.

The functions  $C_i$ and $D_i$ ($i = 0,\ 1,\ 2$) depend on the couplings
of  the Z to  the  leptons  ($v_e,\  a_e,\  v_\tau,\  a_\tau$).  In  the
following  sections we will exploit the  symmetries of the cross section
in order to define  quantities  (in terms of $C$ and $D$) related to the
$\tau$  spin  which can be  measured  accurately  at LEP:  these are the
$\tau$     longitudinal      polarisation      ($\PTAU$)     and     the
transverse--transverse  ($\CTT$) and transverse--normal  ($\CTN$) $\tau$
spin correlations.

\section{$\tau$ LONGITUDINAL POLARISATION}

Due to  the  parity  violation  of  the  neutral  currents,  there  is a
difference  in the  production  cross section (for a given  $\cos\theta$
value) of  the $\tau^-$  with left (L) and right (R)  helicity.  The  $\tau$
polarisation asymmetry is defined as:
\begin{equation}
   \PTAU \ \equiv \ {\sigma_R - \sigma_L \over \sigma_R + \sigma_L}
         \  = \ - \ {D_0 \over C_0}
\end{equation}

\noindent where  $\sigma_{L(R)} \ \equiv \ \sigma\,(e^-  e^+ \rightarrow
\tau^-_{L(R)}\tau^+_{R(L)})$.   Using    equation~(\ref{cross.section}),
$\PTAU$ can be expressed in terms of $\cos\theta$ as:
\begin{equation}
  \PTAU\,(\CT)\ = \ -\ {\AT\,(1+\C2T)\,+\,\AEL\,2\CT
                       \over (1+\C2T)\,+\,\AT\AEL\,2\CT}
\label{ptau}
\end{equation}

\noindent being
\begin{equation}
   \AL \ \equiv \ \RAVL \ = \ {2\,(1-4\STW)\over 1+(1-4\STW)^2}
\label{al}
\end{equation}

\noindent where $\ell = e,\tau$.  A precise  measurement  of
$\AL$  provides  an accurate  determination  of the weak  mixing  angle:
$\Delta\STW \simeq {1\over 8}\, \Delta\AL $.

\subsection{Experimental Determination of $\PTAU$}
\label{sec:ptau}

The kinematics of a $\tau$ decay process  depends on the helicity of the
$\tau$.  The energy  spectrum  observed for a given $\tau$ decay process
($\TAX$) can be expressed in terms of two  contributions,  from left and
right--handed taus respectively:

\begin{equation}
{dN\over dx} = {\sigma_L}\cdot {dN^L\over dx} + {\sigma_R}\cdot {dN^R\over dx}
\label{dndx}
\end{equation}

\noindent  being  $x =  E_X/E_\tau$, the  energy  of the  decay  product
normalised   to   the   $\tau$   energy.  By   simple   spin   arguments
equation~(\ref{dndx}) can be rewritten as:
\begin{equation}
{1\over N} {dN\over dx} = h_0(x) - h_1(x) \cdot \PTAU
\label{ndndx}
\end{equation}

\begin{figure}[htb]
    \mbox{\epsfig{file=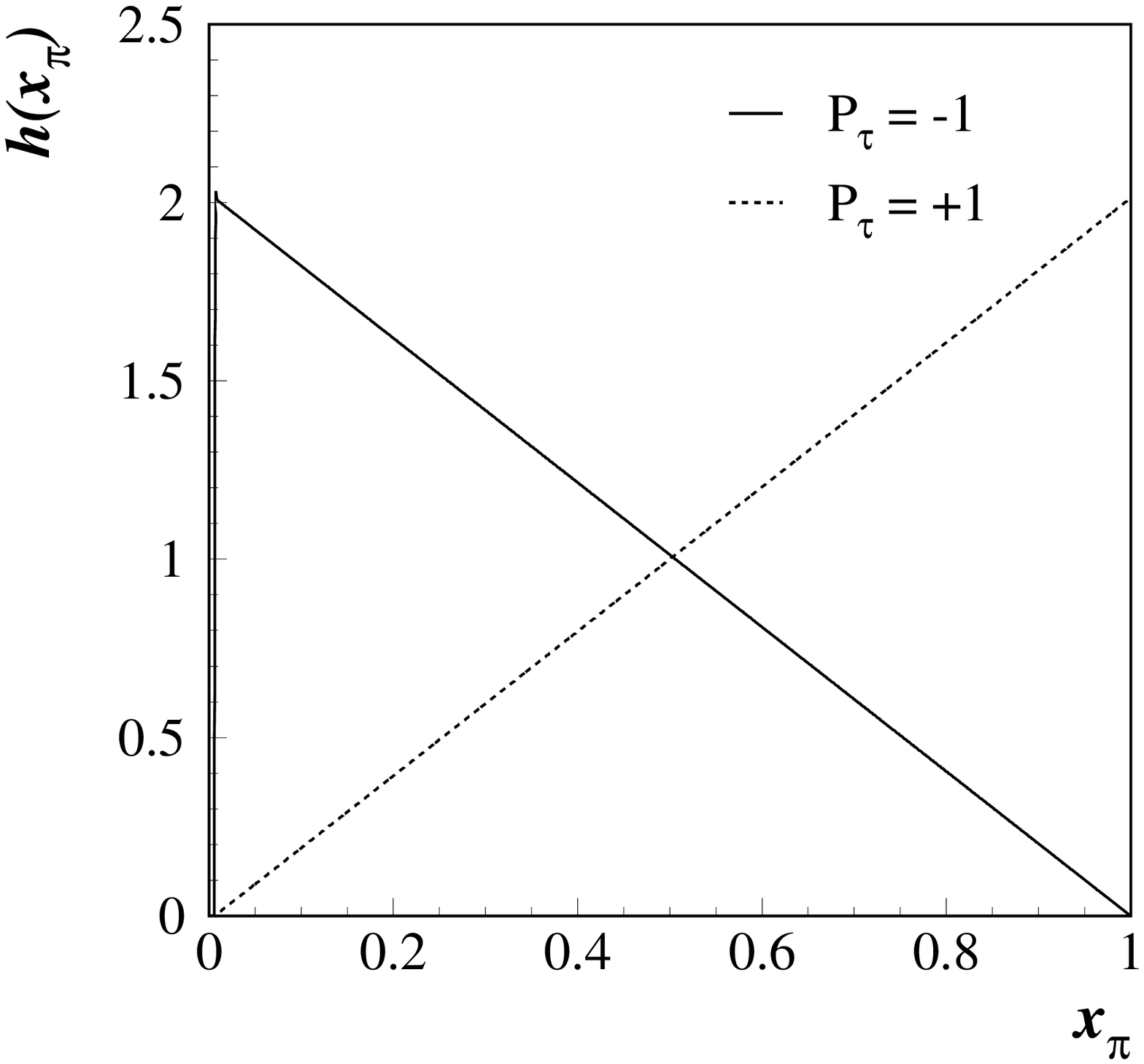,height=3.5cm}
          \epsfig{file=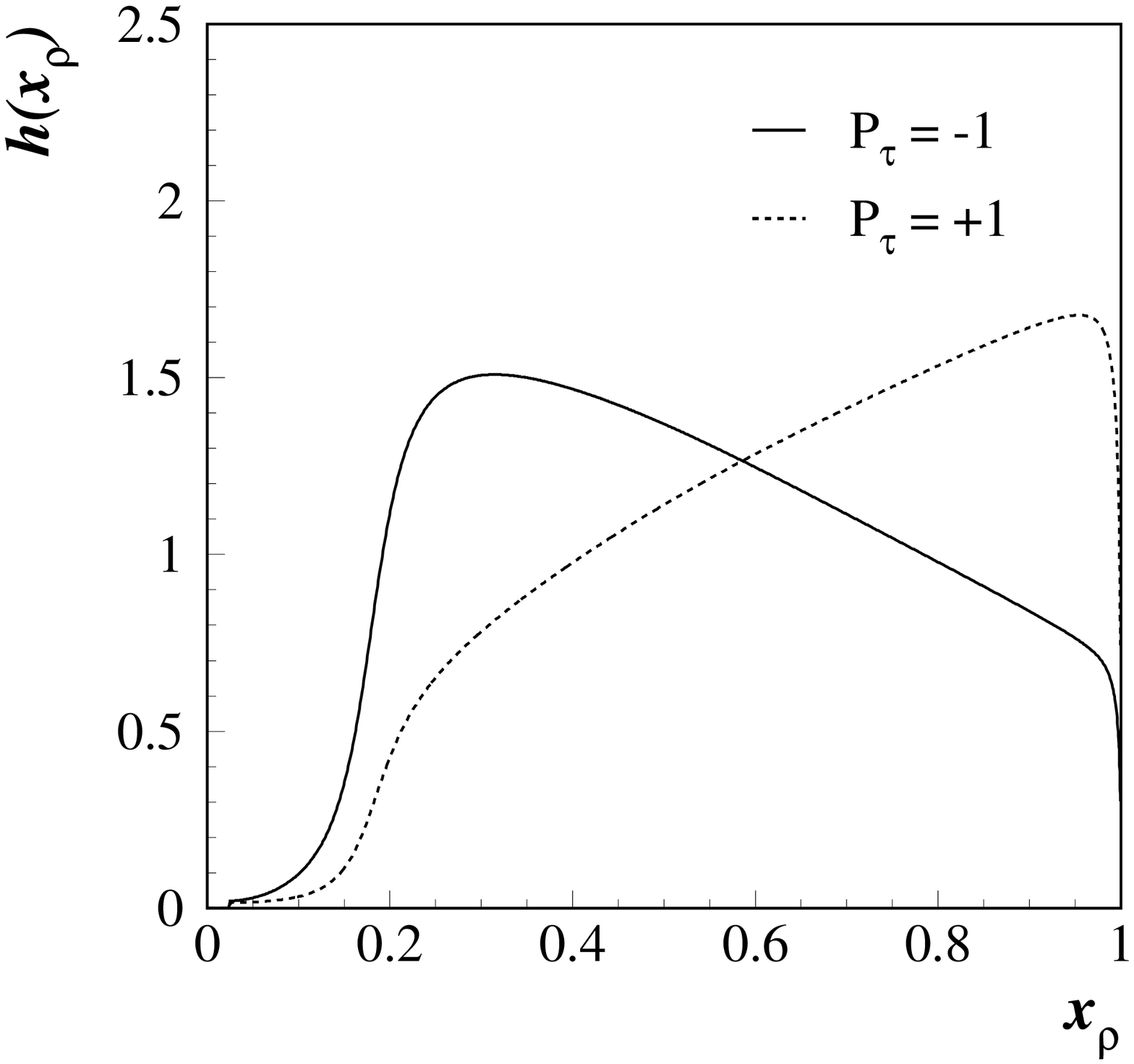,height=3.5cm}}

    \mbox{\epsfig{file=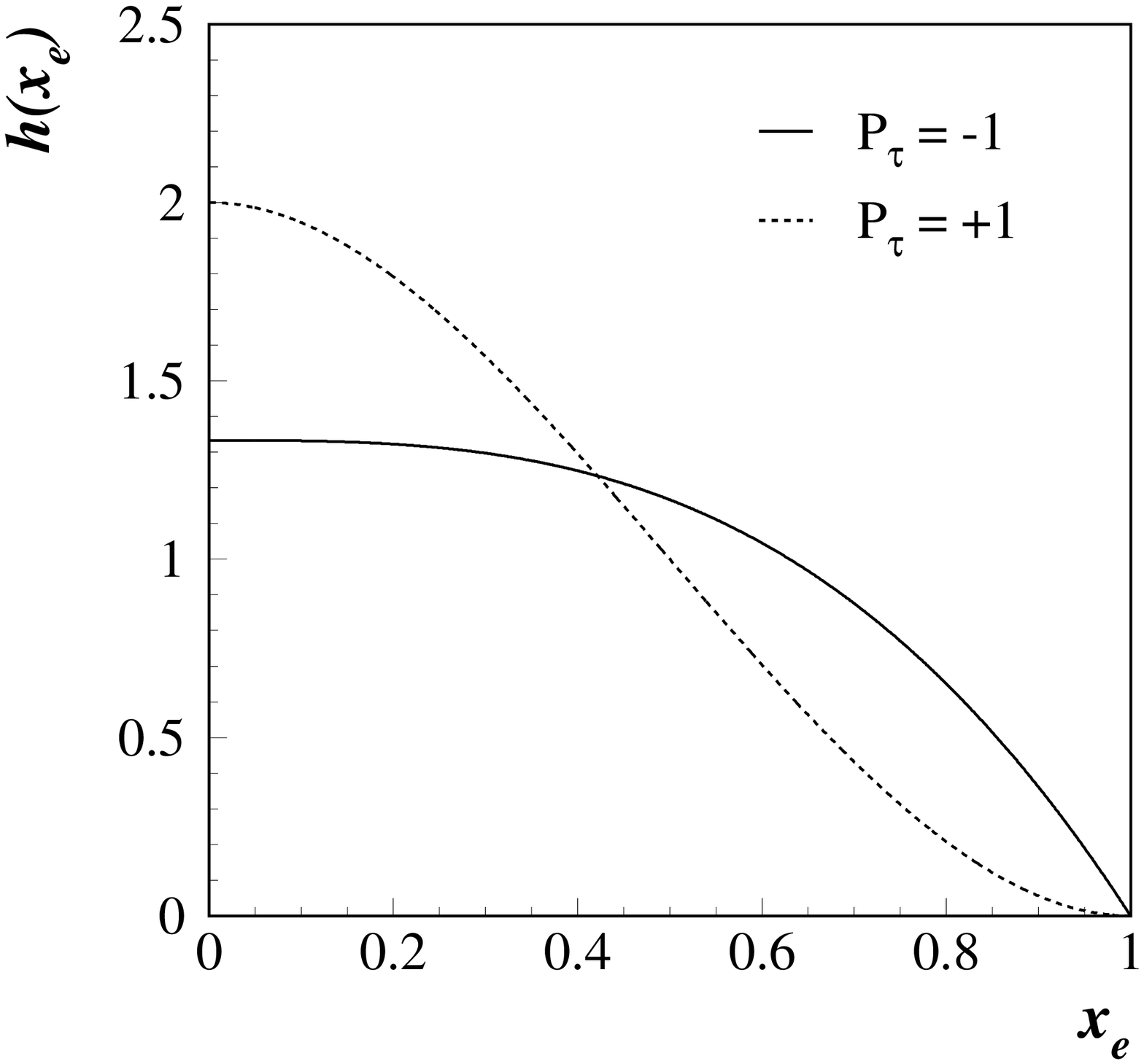,height=3.5cm}
          \epsfig{file=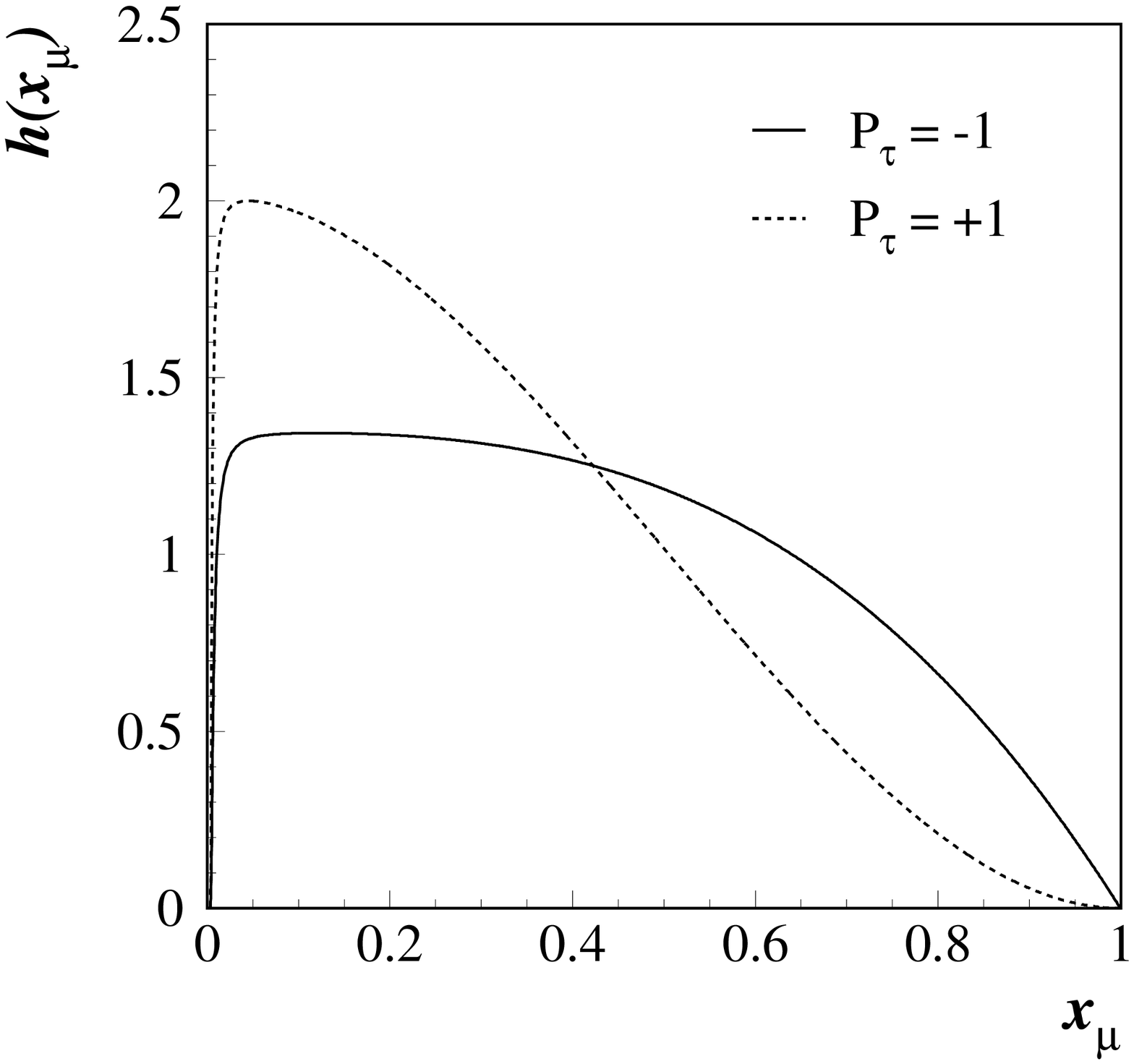,height=3.5cm}}
\vspace*{-0.7cm}
\caption{Energy spectra of the $\tau$ decay products $\pi$, $\rho$, e and
         $\mu$, respectively.}
\label{fig:spectra}
\end{figure}

\noindent The functions  $h_0$ and $h_1$ are different  for every $\tau$
decay  channel  and  depend  mainly  on the  particle  energy,  mass and
spin~\cite{Jad,Hag,Rou}.     A     graphical      representation      of
equation~(\ref{ndndx}), for the two extreme values of $\PTAU$ ($\pm 1$),
is shown in figure~\ref{fig:spectra}~\cite{Pablog}.

The  sensitivity to $\PTAU$ for a given channel  depends on the relative
shape  of  $h_0$  and  $h_1$  and is  maximal  for  the  $\TAP$  channel
(figure~\ref{fig:spectra}).  In the case of multi--pion  $\tau$  decays,
as $\TAR$ and $\TA1$, this sensitivity can be substantially  improved by
defining an optimal variable  ($\omega$)~\cite{Dav}  or set of variables
($\CT^\ast,\  \cos\psi^\ast$)~\cite{Hag} which exploit the kinematics of
the   $\rm    n\pi$    system    in   the    final    state.   Different
experiments~\cite{Opal,L3,Delphi,Aleph}  use different approaches to the
problem.

\begin{figure}[htb]
\vspace{3.4cm}
$$ \mbox{\epsfig{file=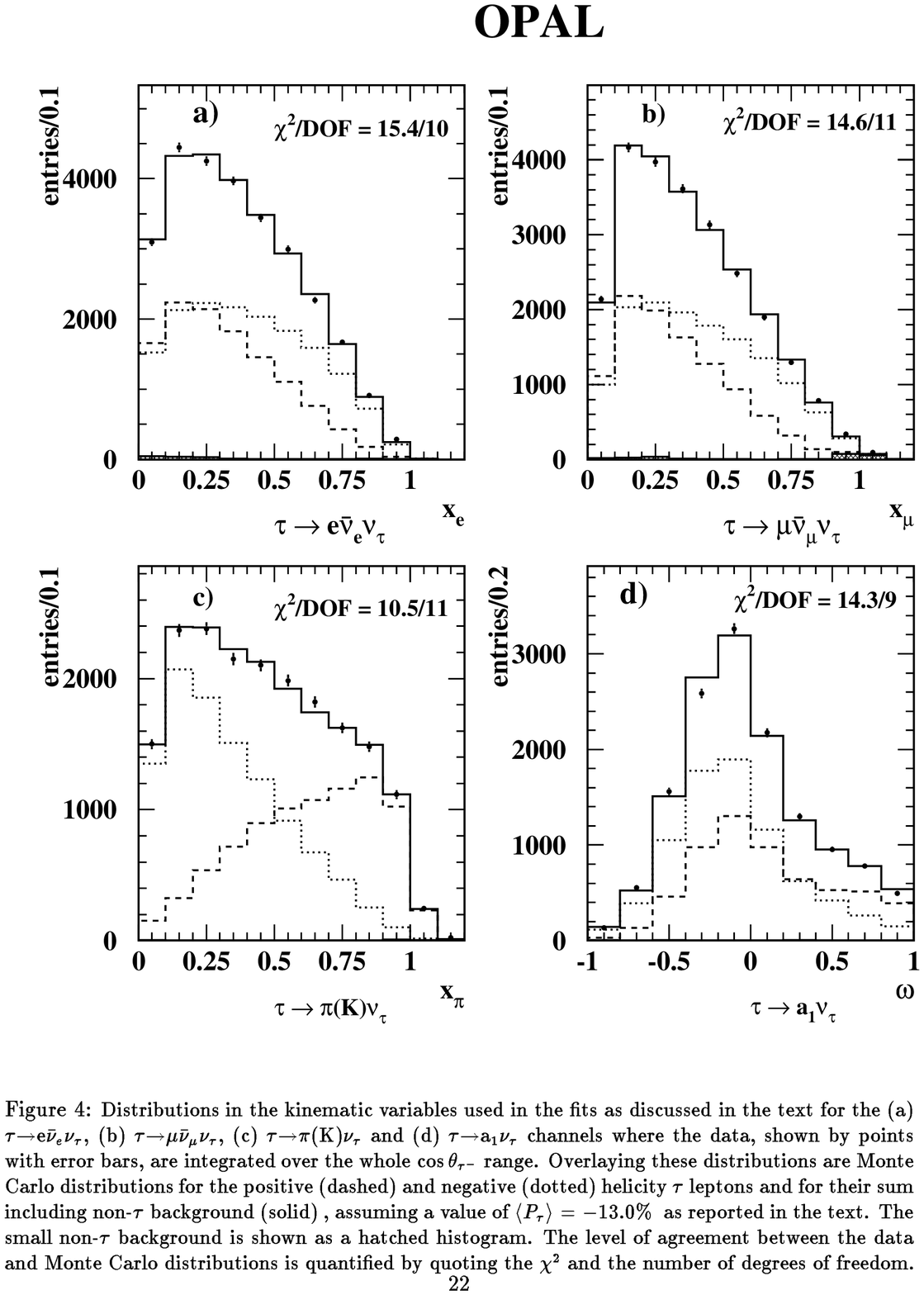,bbllx=65,bblly=265,bburx=545,bbury=454,height=2.8cm,clip=}} $$
\caption{Data spectra (OPAL) for the $\tau$ decay  channels
         e, $\mu$, $\PIK$ and $\A1$. The dots correspond to data
         and the solid line to the fit result. The dotted and dashed lines
         are the contribution from the negative and positive
         helicity spectra.}
\label{fig:OPAL}
\end{figure}

The $\tau$ longitudinal polarisation ($\PTAU$) value is extracted, for a
given $\CT$ range, by performing a fit of the distribution~(\ref{ndndx})
to the data, once the detector and selection  effects have been properly
included  in the  $h_0$  and  $h_1$  functions  (by  using  Monte  Carlo
simulation   and/or   analytical   techniques~\cite{Pablog}).  The  data
spectra  (OPAL~\cite{Opal})  obtained  in the whole  $\CT$  range of the
detector  for the e, $\mu$,  $\PIK$  and  $\A1$  channels  are  shown in
figure~\ref{fig:OPAL}, together with the fit result.

\begin{figure}[htb]
\mbox{\epsfig{file=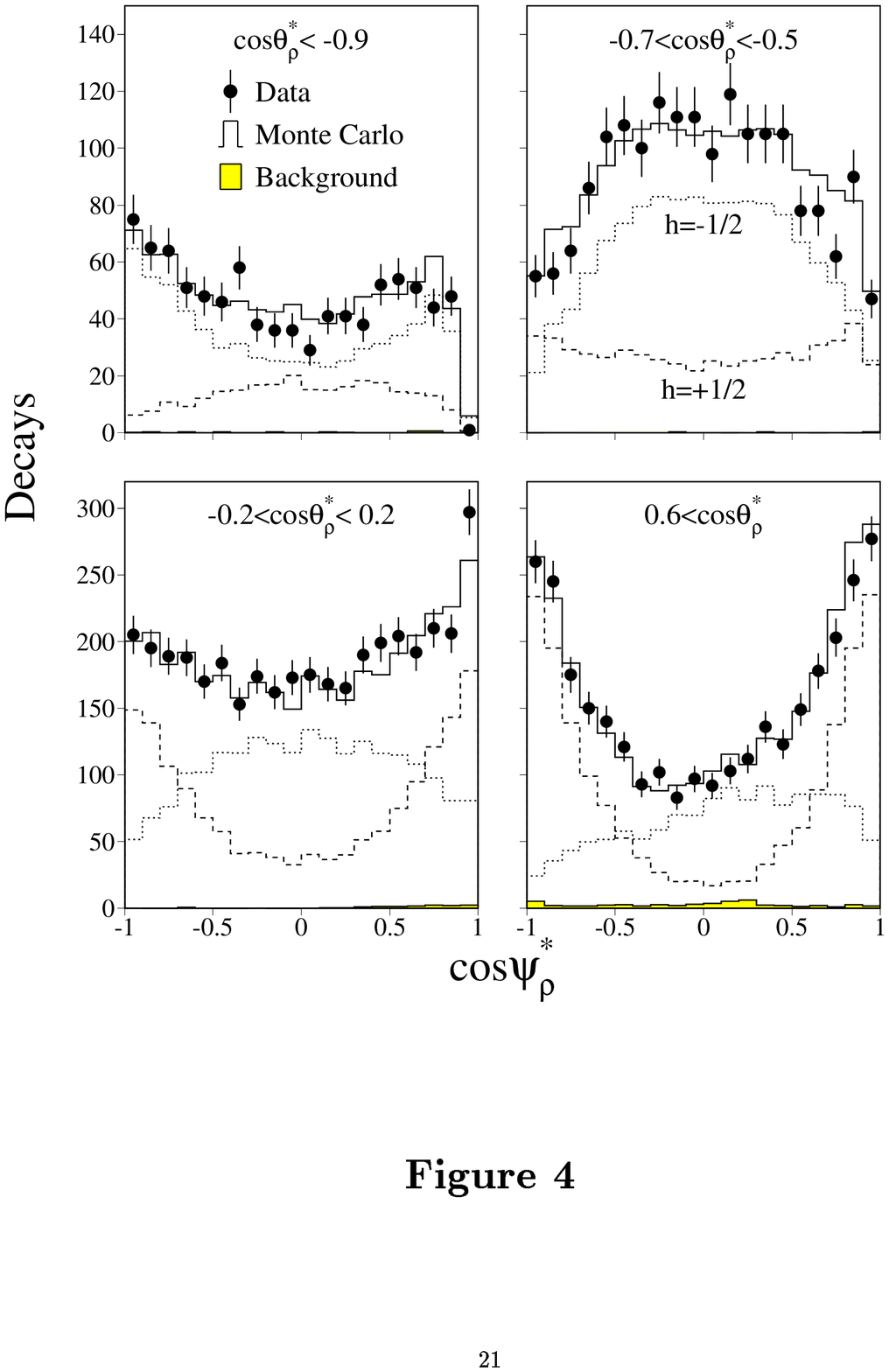,bbllx=50,bblly=235,bburx=495,bbury=740,height=7.5cm,clip=}}
\vspace*{-0.7cm}
\caption{Data spectra from the L3 experiment for the $\tau\rightarrow\rho\nu$
         decay channel. The dots correspond to the data
         and the solid line to the fit result.}
\label{fig:L3}
\end{figure}

As  an  example,  the  two--dimensional  distribution  ($\CT_\rho^\ast$,
$\cos\psi_\rho^\ast$)   for   the  L3   data~\cite{L3}   is   shown   in
figure~\ref{fig:L3}.

\subsection{Data Sample Selection}

The  $\EETT$  processes  are  characterised  for being low  multiplicity
events, with different particle  signatures in opposite  hemispheres (in
most of cases).  In addition,  the  invariant  mass of the event is well
below the Z mass, due to the presence of undetected neutrinos.  The main
background  to this  process is coming  from  $\EEEE$  and $\EEMM$  with
misidentified  particles in the final  state, as well as from two photon
events.

The experiments at LEP are well suited for identifying individual $\tau$
decay channels by considering the tracking chambers  information, showers
from the  electromagnetic  and hadron  calorimeters  and the information
from the muon chambers.  In general the identification  efficiencies are
high, except for multi--pion final states in which $\PI0$ identification
plays a crucial  role.  As a  consequence,  background  in the  hadronic
$\tau$  decay  channels  comes  mainly from  misidentification  of other
hadronic   channels.  In   table~\ref{tab:samples}   we  can   see   the
efficiencies  and  background  contribution  for  the  different  $\tau$
decays.  More details about selection and identification criteria can be
found in \cite{Opal,L3,Delphi,Aleph}.

\begin{table}[hbt]
\setlength{\tabcolsep}{0.5pc}
\newlength{\digitwidth} \settowidth{\digitwidth}{\rm 0}
\catcode`?=\active \def?{\kern\digitwidth}
\caption{Efficiencies (upper line) and background (lower) for the different 
experiments data samples.}
\label{tab:samples}
\vspace{0.3cm}
\begin{tabular}{lcccccc}
\cline{2-7}
            &    e    &    $\mu$  &    $\pi$  &   $\rho$  &   $\A1$   & \#decays \\
\hline\hline
ALEPH       &    59   &      82   &      71   &      59   &      59   &   52  K \\
            &     2   &       4   &       7   &       9   &       9   &         \\
\hline
DELPHI      &    92   &      87   &      60   &      45   &      60   &   71  K \\
            &     5   &       3   &      10   &      15   &      15   &  (49 K) \\
\hline
OPAL        &    96   &      87   &      83   &      70   &      66   &  123  K \\
            &     3   &       2   &      19   &      27   &      25   &         \\
\hline
L3          &    76   &      70   &      72   &      70   &      33   &  111  K \\
            &     5   &       5   &      15   &      11   &      28   &         \\
\hline
\end{tabular}
\end{table}

For the  analysis the DELPHI, OPAL and L3  Collaborations  have used the
data collected between 1990 and 1994, while ALEPH used data collected up
to 1992.  ALEPH and OPAL  results  are final.  Those from  DELPHI and L3
are preliminary.

\subsection{Results from $\PTAU$}

\begin{figure}[htb]
\vspace*{-1.0cm}
$$  \mbox{\epsfig{file=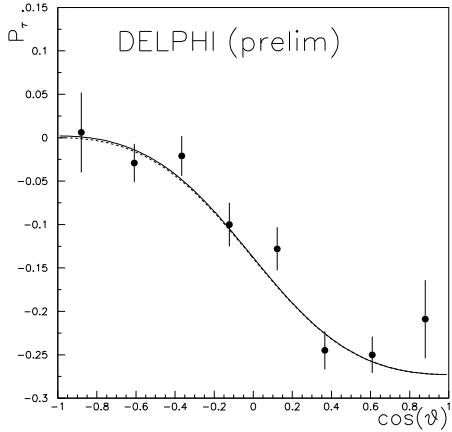,height=6.5cm}} $$
\vspace*{-1.7cm}
\caption{Tau longitudinal polarisation measurement as a function of $\CT$.
         The dots correspond to the data and the solid line to the fit result.}
\label{fig:DELPHI}
\end{figure}

Following the procedure described in section~\ref{sec:ptau},  the $\tau$
longitudinal  polarisation  is measured  for  different  values of $\CT$
(figure~\ref{fig:DELPHI}, DELPHI~\cite{Delphi}).

\begin{figure}[htb]
\vspace{-0.1cm}
$$  \mbox{\epsfig{file=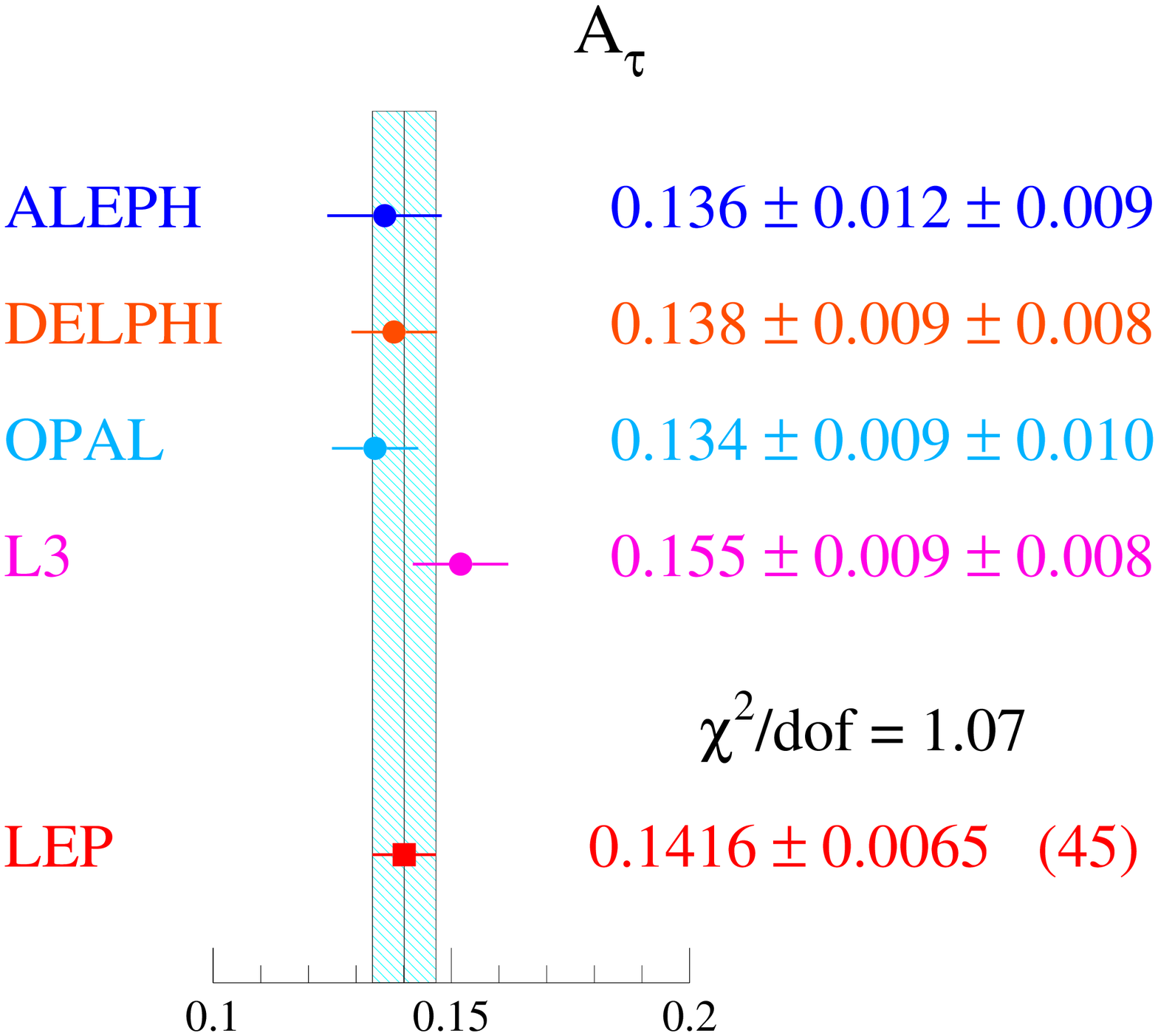,height=5.0cm}} $$
\vspace{-0.7cm}
$$  \mbox{\epsfig{file=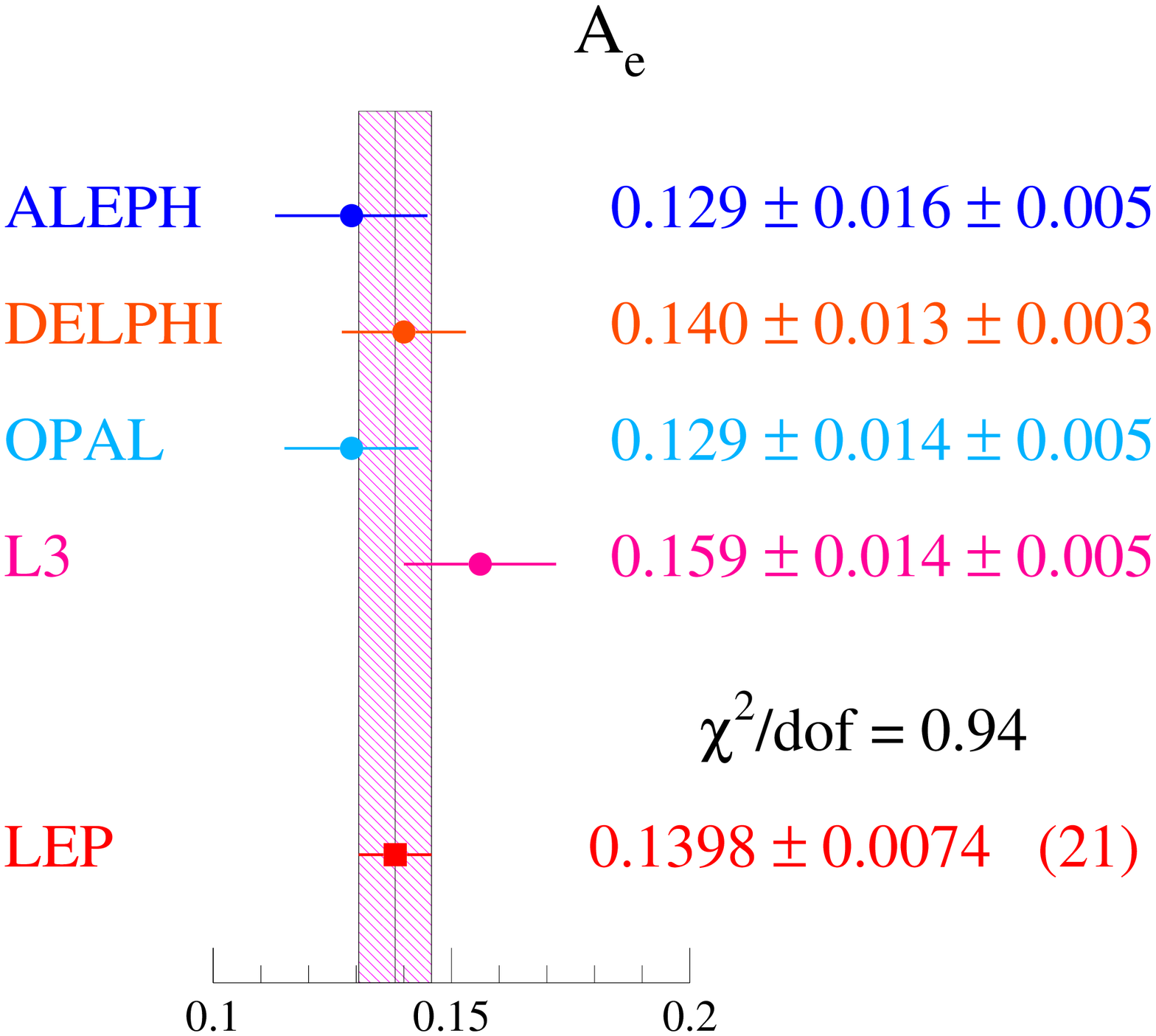,height=5.0cm}} $$
\vspace{-1.6cm}
\caption{LEP results on $\AT$ and $\AEL$.}
\label{fig:results}
\end{figure}

The  values  of  $\AT$  and  $\AEL$   are   obtained   by  fitting   the
expression~(\ref{ptau})   ---which  is  corrected   for  initial   state
radiation,  $\gamma$  exchange and  $\gamma$--Z  interference---  to the
distribution  $\PTAU(\CT)$  in data.  The results of these  measurements
from the four LEP experiments are shown in figure~\ref{fig:results}.

The main sources of systematic  errors come from effects  distorting the
shape of the        spectra,  namely the  calibration  of the  different
detectors and the energy  dependence of the  identification  efficiency,
specially in the channels involving $n\,\PI0$.  

The average LEP result for $\AT$  and  $\AEL$  is:
\begin{eqnarray*}
\qquad   \AT  \ & = & \ 0.1401 \pm 0.0067 \\
\qquad   \AEL \ & = & \ 0.1382 \pm 0.0076
\end{eqnarray*}

\noindent  These results are consistent with the of lepton  universality
of  the  neutral   currents.  Under  this  assumption   $\AU$  has  been
calculated to be:
\begin{eqnarray*}
\qquad \AU \ = \ 0.1393 \pm 0.0050
\end{eqnarray*}

\noindent which gives the following value of the weak mixing angle
(equation~\ref{al}):
\begin{eqnarray*}
\qquad \STW \ = \ 0.2325 \pm 0.0006
\end{eqnarray*}

This result is in agreement with other LEP results~\cite{csh}.

\section{TRANSVERSE AND NORMAL SPIN CORRELATIONS}

The study of spin  correlations  has a special  interest as it  provides
important  additional  tests  of the  Standard  Model.  A non
vanishing  correlation  between the transverse and normal  components of
both taus spins in  $\EETT$  processes  gives rise to terms in the cross
section  in  equation~(\ref{cross.section})  proportional  to $C_2$  and
$D_2$.  The   transverse--transverse   ($\CTT$)  and  transverse--normal
($\CTN$) spin correlations are defined as:
\begin{eqnarray}
\CTT \ \equiv \ \ \  {C_2\over C_0}  \ &  = & \ {|a_\tau|^2-|v_\tau|^2 \over |a_\tau|^2+|v_\tau|^2}        \\
\CTN \ \equiv \ \ \  {D_2\over C_0}  \ &  = & \ -{2\,\Im(v_\tau\,a_\tau^\ast)\over |a_\tau|^2+|v_\tau|^2}
\end{eqnarray}

\noindent  As it  can  be  seen  from  this  definition,  $\CTT$  is not
symmetric in $a_\tau$ and  $v_\tau$:  a value of $\CTT$  different  from
one would point to a Lorentz  structure of the neutral currents which is
different  from the Standard  Model.  In  particular,  a negative  value
would reveal the dominance of the vector  coupling in the neutral $\tau$
current.  Moreover,  $\CTN$ is a CP--odd  observable:  a value of $\CTN$
different  from zero (the SM prediction)  would indicate CP violation in
weak interactions.

\subsection{Experimental Determination}

The determination of the spin  correlations  makes use of the kinematics
of both $\tau$ decay  products in the event.  In this section, some idea
on the  analysis  method will be given.  A detailed  description  of the
different methods and results can be found in~\cite{Fede,Del.c,L3.c}.

Let us consider the process  $\EETT\rightarrow  X_1^- X_2^+ + n \nu$ and
let us define a  reference  system  such  that  $X_1^-$  goes  along the
z--axis and $X_2^+$  lies in the  xz--plane.  If we denote by $\Phi$ the
azimuthal angle of the incident  electron in this reference  system, the
$\Phi $ spectrum for a given final state can be written,  from the cross
section in  equation~(\ref{cross.section}),  as a function of $\CTT$ and
$\CTN$:
\begin{equation}
   \frac{d \sigma}{d\Phi}\ \propto \ \CTT \cos(2\Phi) + \CTN \sin(2\Phi)
\label{ctt.ctn}
\end{equation}

\begin{figure}[htb]
\vspace{-0.3cm}
\mbox{\epsfig{file=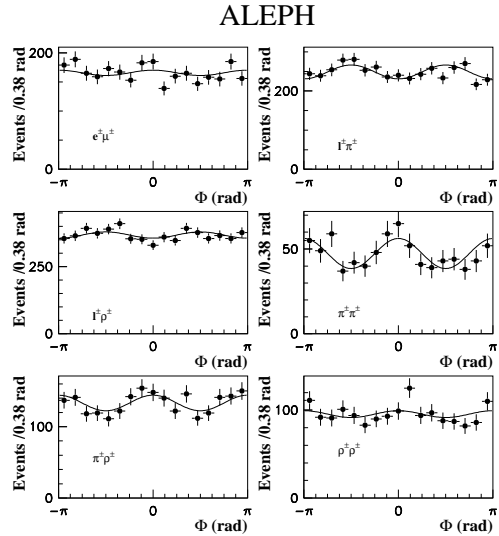,width=7.5cm}}
\vspace{-1.3cm}
\caption{$\Phi$ distribution for the different $\TT$ final states (ALEPH).}
\label{fig:ALEPH}
\end{figure}

The values of $\CTT$ and $\CTN$ are obtained by  performing a likelihood
fit  of  this  function  to  the  $\Phi$   distribution   in  the  data.
Figure~\ref{fig:ALEPH}  shows the $\Phi$ variable spectrum obtained from
data  (ALEPH \cite{Fede})  for the different  $\TT$ final states  (dots)
together with the theoretical  distribution  (line).  The sensitivity to
$\CTT$ and $\CTN$ is higher for final states  involving $\TAP$ and lower
for the leptonic $\tau$ decays.

The main sources of systematic errors are the internal alignment of the
subdetectors and the background due to misidentification of $\TT$ final
states.

\subsection{$\bf\CTT$ and $\bf\CTN$ Results}

The analysis of the spin correlations is based in a data sample of about
37000 $\TT$ events collected in the period  1992--94\footnote{L3  used
only  1994  data}  by the  ALEPH \cite{Fede},  DELPHI~\cite{Del.c}  and
L3~\cite{L3.c}   Collaborations.  

The  results  from  the   corresponding   analyses  is   summarised   in
table~\ref{tab:ctt},  including  the LEP  average.  The  Standard  Model
predictions are in agreement with these results.

\begin{table}[hbt]
\vspace{-0.2cm}
\setlength{\tabcolsep}{0.3pc}
\catcode`?=\active \def?{\kern\digitwidth}
\caption{LEP results on $\CTT$ and $\CTN$.}
\label{tab:ctt}
\begin{tabular}{lcr}
\cline{2-3}
            &       $\CTT$               & \multicolumn{1}{c}{$\CTN$}   \\
\hline\hline
Aleph       & $ 1.00 \pm 0.14 \pm 0.04 $ & $ -0.08 \pm 0.14 \pm 0.02 $  \\
Delphi      & $ 0.87 \pm 0.20 \pm 0.12 $ & \multicolumn{1}{c}{---}      \\
L3          & $ 1.04 \pm 0.26 \pm 0.06 $ & $  0.36 \pm 0.26 \pm 0.05 $  \\
\hline
LEP         & $ 0.98 \pm 0.11          $ & \multicolumn{1}{c}{$0.02 \pm 0.13$}\\
\hline
\end{tabular}
\end{table}

\section*{ACKNOWLEDGEMENTS}

I would  like to thank my  colleagues  from the LEP
collaborations H.  Evans, W.  Lohmann, F.  Matorras, G.
Passaleva,  F.  S\'anchez, H.  Videau, R.  V\"olkert and
M. Wadhwa.

I also wish to thank the German institutions, specially
the DESY--IfH at Zeuthen, for their support to this work.



\end{document}